\documentclass[prl,twocolumn,superscriptaddress,showpacs]{revtex4}

\pdfoutput=1

\usepackage{graphicx}

\usepackage{amsfonts}
\usepackage{amsmath}
\usepackage{amssymb}

\begin{document}

\title{Experimental test of nonlocal realistic theories \\ without the rotational symmetry assumption}

\author{Tomasz Paterek}
\affiliation{Institute for Quantum Optics and Quantum Information (IQOQI),
Austrian Academy of Sciences, Boltzmanngasse 3, A-1090 Vienna, Austria}
\affiliation{The Erwin Schr\"odinger International Institute for Mathematical Physics,
Boltzmanngasse 9, A-1090 Vienna, Austria}

\author{Alessandro Fedrizzi}
\affiliation{Institute for Quantum Optics and Quantum Information (IQOQI),
Austrian Academy of Sciences, Boltzmanngasse 3, A-1090 Vienna, Austria}

\author{Simon Gr\"oblacher}
\affiliation{Institute for Quantum Optics and Quantum Information (IQOQI),
Austrian Academy of Sciences, Boltzmanngasse 3, A-1090 Vienna, Austria}

\author{Thomas Jennewein}
\affiliation{Institute for Quantum Optics and Quantum Information (IQOQI),
Austrian Academy of Sciences, Boltzmanngasse 3, A-1090 Vienna, Austria}

\author{Marek \.Zukowski}
\affiliation{Faculty of Physics, University of Vienna, Boltzmanngasse 5, A-1090 Vienna, Austria}
\affiliation{Institute of Theoretical Physics and Astrophysics,
University of Gda\'nsk, ul. Wita Stwosza 57, PL-08-952 Gda\'nsk, Poland}

\author{Markus Aspelmeyer}
\affiliation{Institute for Quantum Optics and Quantum Information (IQOQI),
Austrian Academy of Sciences, Boltzmanngasse 3, A-1090 Vienna, Austria}

\author{Anton Zeilinger}
\affiliation{Institute for Quantum Optics and Quantum Information (IQOQI),
Austrian Academy of Sciences, Boltzmanngasse 3, A-1090 Vienna, Austria}
\affiliation{Faculty of Physics, University of Vienna, Boltzmanngasse 5, A-1090 Vienna, Austria}

\begin{abstract}
We analyze the class of nonlocal realistic theories that was originally considered by Leggett [Found. Phys. \textbf{33}, 1469 (2003)] and tested by us in a recent experiment [Nature \textbf{446}, 871 (2007)]. We derive an incompatibility theorem that works for finite numbers of polarizer settings and that does not require the previously assumed rotational symmetry of the two-particle correlation functions. The experimentally measured case involves seven different measurement settings. Using polarization-entangled photon pairs, we exclude this broader class of nonlocal realistic models by experimentally violating a new Leggett-type inequality by 80 standard deviations \cite{PRL}.
\end{abstract}

\date{\today}


\maketitle

It is the essence of Bell's theorem~\cite{Bell1964,CHSH} and its many experimental tests, for example~\cite{Freedman,Aspect,Weihs,Rowe}, that no theory based on the joint assumptions of locality and realism can serve as an alternative underlying explanation of quantum phenomena (further relevant references can be found in \cite{Leggett2003, Groeblacher2007}). Recently, a new type of incompatibility theorem was introduced by Leggett~\cite{Leggett2003} that allowed for the first time a test of a specific and intuitive class of nonlocal realistic hidden-variable theories. The experimental exclusion of this class has been reported in Ref.~\cite{Groeblacher2007}. 
Since the original inequality required infinitely many measurement settings, it was necessary to supplement it by an assumption of rotational symmetry of the correlation functions in each measurement plane. 
The main point of the present paper is that this assumption is no longer needed.
We prove a theorem which allows an experimental test of a broader class of nonlocal realistic theories. In contrast to the original derivation~\cite{Leggett2003,Groeblacher2007}, the correlation functions of measurement outcomes are no longer constrained to be rotationally symmetric. We test the resulting new inequality by using a high-efficiency, high-fidelity source of polarization-entangled photon pairs~\cite{FEDRIZZI07}.

\emph{Generalized incompatibility theorem.}---The theories under investigation aim to uphold realism, while allowing for nonlocal influences.
It is additionally assumed that particles of a defined property behave locally
in concordance with quantum laws.
Specifically, we consider polarization
measurements on pairs of photons. Additionally to realism -- the
assumption that measurement outcomes are well defined prior to and
independent of the measurements-- the theories attribute polarization to
single photons. Particles with the same polarization build up subensembles
in which Malus' law is assumed to hold. All observationally accessible
expectation values are computed as statistical mixtures over such subensembles.
In particular, the theories provide a model for all
experiments in which a Clauser-Horne-Shimony-Holt inequality \cite{CHSH} is violated 
and they reproduce all perfect correlations of the Bell singlet state 
\footnote{Note that such nonlocal realistic models do not follow the Furry-Schr\"odinger hypothesis \cite{Furry1936}.}.

Let us introduce the notation, which follows Ref.~\cite{Groeblacher2007}.
Alice and Bob perform measurements on individual photons of the pairs. All observables are parameterized by vectors on the Poincar{\'e} sphere. Alice (Bob) performs measurements along direction $\vec a_k$ ($\vec b_l$). The corresponding correlation function is denoted by $E_{kl}(\xi,\varphi)$, where the angles $\xi$ and $\varphi$ parameterize the position of the vectors $\vec a_k$ and $\vec b_l$ within the plane spanned by them ($\varphi$ is the angle between $\vec a_k$ and $\vec b_l$, and $\xi$ describes the orientation of the vector bisecting the angle $\varphi$).

The inequality of the original derivation makes use of the averaged correlation functions
\begin{equation}
\overline{E}_{kl}(\varphi) = \frac{1}{2 \pi} \int_0^{2 \pi}  E_{kl}(\xi,\varphi) d \xi.
\label{AVG_CORR}
\end{equation} 
In order to get well approximated values of the averaged correlation functions,
we would have to perform a large number of measurements.
Following the earliest experimental tests of local realism \cite{Freedman, Clauser1976,Fry1976},
we resorted to the assumption that the observed correlation functions must have the property
of rotational symmetry, i.e.,
that the correlation functions depend only on the angle $\varphi$.
This is experimentally well-established.
Various checks of this symmetry were performed, e.g., in Refs.~\cite{Aspect1981,Shih1994,Kwiat1995,Weihs,Aspect2002},
 \footnote{Note that this does not imply a pre-assumption of the validity of quantum theory,
because only assumptions about the phenomenologically defined quantities
``correlation functions'' are made.}.

The basic idea of the new incompatibility theorem is to replace the averaged correlations (\ref{AVG_CORR})
by the finite sum of non-averaged correlation functions~\cite{PhDThesis_Paterek}. A detailed derivation of the set of new inequalities, that essentially parallels the one in ~\cite{Groeblacher2007}, is provided in the Appendix. In the experimentally most easily realizable case, this leads to the following inequality:
\begin{eqnarray}
S &\equiv& \left| E_{kl}(0,\varphi) + E_{k'l'}(\frac{\pi}{2},\varphi)+E_{mn}(0, 0) + E_{m'n'}(\frac{\pi}{2},0) \right| \nonumber \\
&+&
\left| E_{qp}^{\perp}(0,\varphi) + E_{q'p'}^{\perp}(\frac{\pi}{2},\varphi)+E_{rs}^{\perp}(0, 0) + E_{r's'}^{\perp}(\frac{\pi}{2}, 0) \right| \nonumber \\
& \le & 8 - 2 |\sin \frac{\varphi}{2}|.
\label{ROT_INV_INEQ}
\end{eqnarray}
It is a necessary consequence of the derivation that settings of the correlation functions in the second modulus have to lie in any plane orthogonal to that defined by the settings of correlation functions in the first modulus. Thus, they have an additional superscript $\perp$.

Again, quantum predictions violate the new inequality (\ref{ROT_INV_INEQ}). For example, the two-particle singlet state yields the quantum correlation function $E(\varphi) = - \cos \varphi$. For this state, $S = 4|1+\cos\varphi|$. Maximum violation is obtained for angle $\varphi_{\max} \approx 14.6^{\circ}$ at which the bound equals $7.746$, in contrast to the quantum value of $7.871$ at the left-hand side of the inequality.
The ratio of the bound of Eq.~(\ref{ROT_INV_INEQ}) for $\varphi_{\max}$ 
and the quantum value is $0.984$.
As a consequence, the minimal visibility of the two-particle interference
that is required to unambiguously observe a violation at the optimal difference angle must be larger than $98.4 \%$.

\begin{figure}
\begin{center}
\includegraphics[scale=0.75]{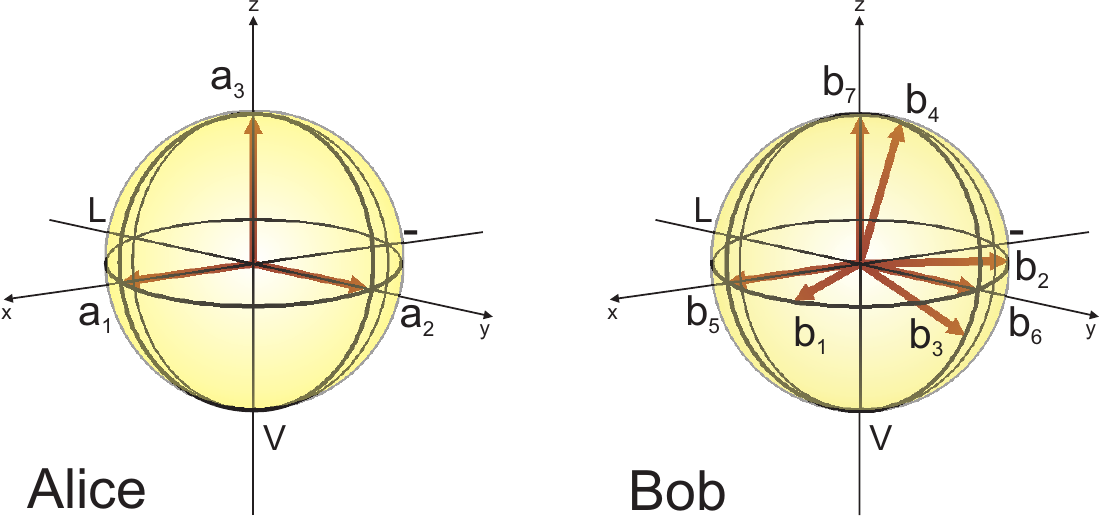}
\end{center}
\caption{The Poincar{\'e} vectors corresponding to measurement settings of Alice and Bob for the maximal violation of inequality (\ref{ROT_INV_INEQ}).}
\label{ROT_INV_FREE_FIG}
\end{figure}

In order to test the inequality, it is sufficient that Alice and Bob choose among seven different pairs of settings.
Alice's setting vectors are
\begin{equation}
\vec a_1 = (1,0,0), \quad
\vec a_2 = (0,1,0), \quad
\vec a_3 = (0,0,1).
\label{SETTINGS_A}
\end{equation}
Bob's setting vectors are
\begin{eqnarray}
\vec b_1 & = & (\cos\varphi_{\max},\sin\varphi_{\max},0), \nonumber \\
\vec b_2 & = & (-\sin\varphi_{\max},\cos\varphi_{\max},0), \nonumber \\
\vec b_3 & = & (0,\cos\varphi_{\max},-\sin\varphi_{\max}), \nonumber \\
\vec b_4 & = & (0,\sin\varphi_{\max},\cos\varphi_{\max}), \nonumber \\
\vec b_5 & = & \vec a_1, \nonumber \\
\vec b_6 & = & \vec a_2, \nonumber \\
\vec b_7 & = & \vec a_3.
\label{SETTINGS_B}
\end{eqnarray}
All the vectors are depicted on the Poincar{\'e} sphere
in Fig.\ \ref{ROT_INV_FREE_FIG}.
For these settings, inequality (\ref{ROT_INV_INEQ}) reads:
\begin{eqnarray}
&& |E_{11} + E_{22} + E_{15} + E_{26}|
+ |E_{23} + E_{34} + E_{26} + E_{37}| \nonumber \\
&& \le  8 - 2 |\sin\frac{\varphi_{\max}}{2}|=7.746.
\label{MAX_INEQ}
\end{eqnarray}

\begin{figure}[b]
\centering
\includegraphics[width=6cm]{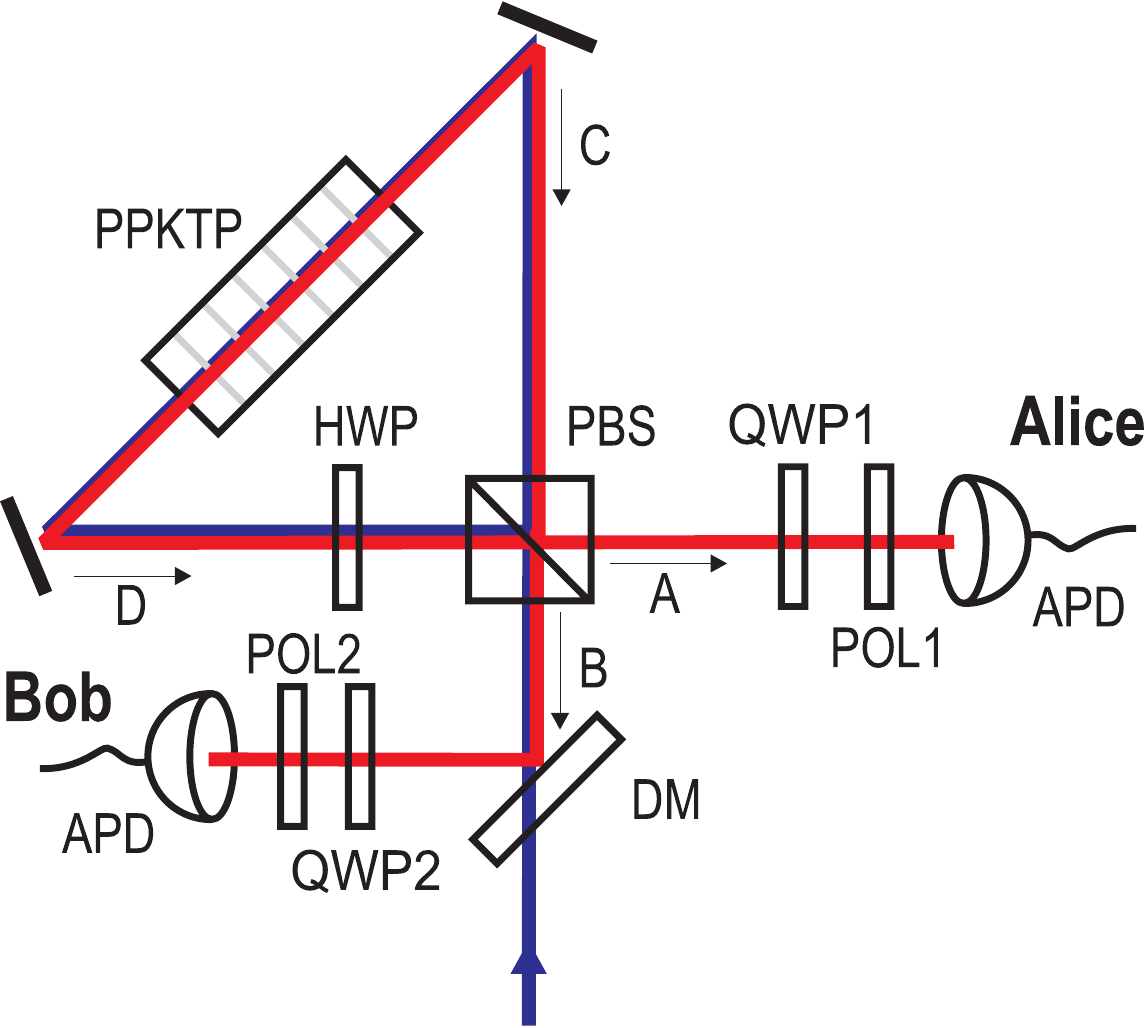}
\caption{Scheme of the experiment: A 405 nm diode laser is focused into a
PPKTP crystal inside a polarization Sagnac loop comprised of a
dual-wavelength polarizing beam splitter (PBS), a dual-wavelength
half-wave plate (HWP), and two laser mirrors. In the crystal,
orthogonally polarized photon pairs are created in modes C and
D. At the PBS, these photon pairs are coherently superposed, the horizontally polarized
photons are transmitted, and the vertically polarized ones are reflected such that the
resulting two-photon state in modes A and B is polarization
entangled. The polarization states of the photons are analyzed and
measured by a combination of a quarter-wave plate (QWP), a polarizer
(POL), and an Silicon avalanche photodiode (APD).}
\label{fig:setup}
\end{figure}

\emph{Experiment.}---We tested  inequality (\ref{MAX_INEQ}) by measuring polarization correlations on polarization-entangled photon pairs for the 
settings (\ref{SETTINGS_A}) and (\ref{SETTINGS_B}) with $\varphi_{\max} \approx 14.6^{\circ}$. We used a high-efficiency, high-fidelity pair source based on spontaneous parametric downconversion (SPDC) in periodically poled KTiOPO$_4$ (PPKTP) inside a polarization Sagnac interferometer (Fig.~\ref{fig:setup}).
This configuration was originally demonstrated in Ref.~\cite{KIM06};
the setup used here is explained in more detail in Ref.~\cite{FEDRIZZI07}. 
In our experiment, the source was aligned to emit the
singlet state $| \psi^- \rangle = \frac{1}{\sqrt{2}} \left( | H \rangle_A | V \rangle_B - | V \rangle_A | H \rangle_B \right)$, where $| H(V) \rangle_{A(B)}$ denotes a horizontally (vertically) polarized photon that leaves the interferometer towards Alice (Bob).

Polarization measurements were performed by passing each photon of a pair through a combination of quarter-wave plates and polarizers and by subsequently detecting it by single-photon avalanche photodiodes. In that way, arbitrary setting directions on the Poincar{\' e} sphere can be realized (see, e.g.,~\cite{Walther2005}). Joint detection events between Alice and Bob were registered within a coincidence time window of 4.4~ns by analyzing the individual events via a field-programmable gate array (FPGA). 
Without polarizers in place, we typically observed coincidence count-rates of $\sim 20000$ per second with single count rates of $\sim 80000$ (the same for Alice and Bob) at an optical pump power of 0.6~mW. 
Polarization correlation measurements along directions $|H/V\rangle$ (corresponding to correlation function $E_{37}$), $|\pm\rangle=\frac{1}{\sqrt{2}}(|H\rangle\pm|V\rangle)$ ($E_{15}$), and $|R/L\rangle=\frac{1}{\sqrt{2}}(|H\rangle\pm i|V\rangle)$ ($E_{26}$) reveal an actual visibility of $(99.70\pm0.02)\%$, $(99.47\pm0.02)\%$, and $(99.25\pm0.03)\%$.
The average two-photon interference visibility of $(99.47\pm0.01)\%$
clearly  surpasses the required limit of $98.4\%$.

We measured the expectation values of inequality (\ref{MAX_INEQ}) by using the optimal settings described above. The resulting value of its left-hand side, S, is compared to the bound of 7.746 valid for the generalized class of nonlocal realistic models. The result is summarized in Tab.~\ref{tab:results}.
\begin{table}[t]
\caption{Comparison of ideal quantum theoretical and experimental
expectation values, $E_{ij}$, that enter inequality (\ref{MAX_INEQ}).
Below, the theoretical $S$-value [the left-hand side of (\ref{MAX_INEQ})]
is compared with the experimental one.
The discrepancy between theory and experiment is explained by taking into account an average two-photon interference visibility of $(99.47\pm0.01)\%$ and the inaccuracy in setting the local measurement direction of approximately $\pm0.5^{\circ}$. The standard deviations $\sigma_E$ and $\sigma_S$ are the relevant errors for calculating the violation (see text) and are essentially due to Poissonian count statistics of the measured coincidence rates.}
\begin{tabular}{lclclcl}
\hline \hline
 & & $E_{\mathrm{theory}}$ & & $E_{\mathrm{experiment}}$ & & $\sigma_E$ \\
\hline
$E_{11}$ & \hspace{0.7 cm} &  -0.9677 & \hspace{0.7 cm} & -0.9749 & \hspace{0.7 cm} & 0.0005 \\
$E_{22}$ & & -0.9677 & & -0.9733 & & 0.0005 \\
$E_{15}$ & & -1 & & -0.9947 & & 0.0002 \\
$E_{26}$ & & -1 & & -0.9925 & & 0.0003 \\
$E_{23}$ & & -0.9677 & & -0.9601 & & 0.0007 \\
$E_{34}$ & & -0.9677 & & -0.9662 & & 0.0006 \\
$E_{37}$ & & -1 & & -0.9970 & & 0.0002 \\
& & $S_{\mathrm{theory}}$ & & $S_{\mathrm{experiment}}$ & & $\sigma_S$ \\
& & 7.8708 & & 7.8511 & & 0.0013\\
 \hline \hline
\end{tabular}
\label{tab:results}
\end{table}
Error analysis is performed by taking into account both Poissonian counting statistics and the inaccuracy in setting the measurement direction with the quarter-wave plates and polarizers. 
In summary, we observe a violation of inequality (\ref{MAX_INEQ}), as given by the minimum distance between the measured $S$-value and the theoretical bound, by 80 standard deviations.

\emph{Conclusions}.---Based on the recent works by Leggett \cite{Leggett2003} and by us \cite{Groeblacher2007},
we derive a new incompatibility theorem
that does not require the previously assumed rotational
symmetry of the two-particle correlation functions and
hence puts to test a broader class of nonlocal realistic
hidden-variable theories. We demonstrate an experimental
violation of the resulting new inequality by 80 standard
deviations.

We acknowledge discussions with {\v C}. Brukner and A. J. Leggett. The work was supported by the Austrian Science Fund (FWF), by the Austrian Research Promotion Agency (FFG), by the EC funded project QAP and by the Foundational Questions Institute (FQXi).

\emph{Note added}.---Recently, we were informed of a closely related work by Branciard \emph{et al}. \cite{Scarani2007}.

\emph{Appendix.}---We derive an infinite set of inequalities,
the simplest case of which is the inequality (\ref{ROT_INV_INEQ}).
We follow exactly all the steps as in the Supplementary Information (SI) of Ref. \cite{Groeblacher2007} up to the formula (27). In an abbreviated notation of the integration, which utilizes the fact that explicit integration boundaries play no role in the derivation, it reads (for notation and symbols see \cite{Groeblacher2007}):
\begin{widetext}
\begin{equation}
E_{kl}(\xi_{kl}, \varphi_{kl}) \le 1 - 2
\int d \vec u d \vec v F(\vec u, \vec v)
\sqrt{n_2^2
\cos^2(\frac{\varphi_{kl} - \chi_{uv}}{2}) + n_1^2 \sin^2(\frac{\varphi_{kl} -
\chi_{uv}}{2})} |\cos(\xi_{kl} - \psi_{uv} + \alpha)|.
\label{XI_PHI_CORR}
\end{equation}
In the original proof, Eq.~(6) was averaged over
all possible setting angles in one plane. 
Here, we avoid this entirely by considering ${N}\geq2$ correlation functions for settings from a {\em single plane}:
$E_{k^{n}l^{n}}(\xi_{k^nl^n} = n\pi/N, \varphi_{k^{n}l^{n}})$, with $n=0,1,2,...,N-1$.
We set the angle between the setting vectors
which enter all correlation functions to be the same, i.e.
$\varphi \equiv \varphi_{k^{n}l^{n}}$.
The settings of the $n$th correlation function
are rotated by $\frac{\pi}{N}$
with respect to the settings of the $(n-1)$th correlation function.
We sum up inequalities (\ref{XI_PHI_CORR}) for all $N$ correlation functions:
\begin{equation}
\sum_{n=0}^{N-1}E_{k^nl^n}(\frac{n\pi}{N}, \varphi) \le N - 2
\int d \vec u d \vec v F(\vec u, \vec v)
\sqrt{n_2^2
\cos^2(\frac{\varphi - \chi_{uv}}{2}) + n_1^2 \sin^2(\frac{\varphi -
\chi_{uv}}{2})}\sum_{n=0}^{N-1} |\cos(n\frac{\pi}{N} - \psi_{uv} + \alpha)| .
\end{equation}
We utilize the following inequality
\begin{equation}
\sum_{n=0}^{N-1} |\cos(n\pi/N- \psi_{uv} + \alpha)|\geq \cot\frac{\pi}{2N},
\end{equation}
and obtain
\begin{equation}
\frac{1}{N}\sum_{n=0}^{N-1}E_{k^nl^n}(\frac{n\pi}{N}, \varphi)
 \le 1 - 2\frac{K(N)}{N}\int d \vec u d \vec v F(\vec u, \vec v)
\sqrt{n_2^2
\cos^2(\frac{\varphi - \chi_{uv}}{2}) + n_1^2 \sin^2(\frac{\varphi -
\chi_{uv}}{2})},  \label{INEQ-N}
\end{equation}
where
$K(N)=\cot\frac{\pi}{2N}$.
This inequality is valid for any choice of observables within a {\em single} plane.
We introduce $N$ new observable vector pairs in this plane,
again rotated by $n \frac{\pi}{N}$ with respect to an initial pair.
The correlation functions of these
new observables will be denoted by
$E_{i^nj^n}(\xi_{i^nj^n}'=\frac{n\pi}{N},\varphi_{i^nj^n}')$,
where the angles $\xi_{i^nj^n}'$ can be relative to another axis than $\xi_{k^nl^n}$.
Again, we set $\varphi_{i^nj^n}' = \varphi'$
and arrive at the analog of inequality (\ref{INEQ-N}).
The sum of these inequalities reads
\begin{eqnarray}
&& \frac{1}{N}\left
(\sum_{n=0}^{N-1}E_{k^nl^n}(\frac{n\pi}{N}, \varphi)+\sum_{n=0}^{N-1}E_{i^nj^n}(\frac{n\pi}{N}, \varphi')\right)
 \nonumber \\
&\le& 2- 2\frac{K(N)}{N} \int d \vec u d \vec v F(\vec u, \vec v)
\left( \sqrt{n_2^2 \cos^2 \frac{\varphi - \chi_{uv}}{2} + n_1^2
\sin^2 \frac{\varphi - \chi_{uv}}{2}} +\sqrt{n_2^2 \cos^2
\frac{\varphi' - \chi_{uv}}{2} + n_1^2 \sin^2 \frac{\varphi' -
\chi_{uv}}{2}} \right).
\label{PLANE_INEQ}
\end{eqnarray}
To estimate the bound, we use the manipulation involving the triangle inequality, which follows exactly relations (31)-(33) of the SI, to get 
\begin{eqnarray}
\frac{1}{N} \left| \sum_{n=0}^{N-1}E_{k^nl^n}(\frac{n\pi}{N}, \varphi)+\sum_{n=0}^{N-1}E_{i^nj^n}(\frac{n\pi}{N}, \varphi') \right| & \le & 
  2
- \sqrt{2}\frac{K(N)}{N} |\sin\frac{\varphi - \varphi'}{2}|  \int d \vec u d \vec v F(\vec u, \vec v) \sqrt{\sin^2{\theta_u} + \sin^2{\theta_v}} .
\label{XY}
\end{eqnarray}
This formula replaces (35) of the SI. As we see, the net change on the right hand side is that $\frac{2\sqrt{2}}{\pi}$, which is equal to $\lim_{N\rightarrow\infty}\sqrt{2}\frac{K(N)}{N}$, is replaced by $\sqrt{2}\frac{K(N)}{N}$. For settings within a plane {\em orthogonal} to the initial one, we get a similar inequality. 
After adding the inequalities for the two orthogonal planes
we set angles $\varphi$ in those planes to be equal, and angles $\varphi'$ to zero.
Next, we utilize the fact $\sqrt{\sin^2{\theta_u} + \sin^2{\theta_v}}+\sqrt{\sin^2{\theta'_u} + \sin^2{\theta'_v}}\geq \sqrt{2}$, see (37)-(43) of the SI,
which leads us to  the following  set of inequalities:
\begin{equation}
\frac{1}{N} \left| \sum_{n=0}^{N-1}E_{k^nl^n}(\frac{n\pi}{N}, \varphi)
+\sum_{n=0}^{N-1}E_{i^nj^n}(\frac{n\pi}{N}, 0)\right|
+ \frac{1}{N} \left| \sum_{n=0}^{N-1}E^\bot_{q^np^n}(\frac{n\pi}{N}, \varphi)
+\sum_{n=0}^{N-1}E^\bot_{r^ns^n}(\frac{n\pi}{N},0)\right|
\leq  4 - 2 \frac{K(N)}{N} |\sin{\frac{\varphi}{2}}|.
\end{equation}
\end{widetext}
In the limit $N \to \infty $, since $\lim_{N \to \infty} 2 \frac{K(N)}{N} = 4/\pi$,
we recover the inequality published in \cite{Groeblacher2007}. The simplest one, for $N=2$, has the explicit form of inequality (\ref{ROT_INV_INEQ}) and uses only seven pairs of measurement settings.

\end{document}